# A 1.16-μm-radius disk cavity in a sunflower-type circular photonic crystal with ultrahigh quality factor

Xufeng Zhang, Xiankai Sun, and Hong X. Tang*

*Department of Electrical Engineering, Yale University, 15 Prospect St., New Haven, CT 06511, USA*

*\* Corresponding author: hong.tang@yale.edu*

We present a 1.16-μm-radius disk cavity with ultrahigh quality ($Q$) factor by embedding the disk into a sunflower-type circular photonic crystal (CPC). The band gap of the CPC reduces the bending loss of the whispering-gallery mode of the disk, leading to a simulated $Q$ of $10^7$, at least one order of magnitude higher than a bare disk of the same size. The design is experimentally verified with a record high loaded $Q$ of $7.4 \times 10^5$ measured from an optimized device fabricated on a silicon-on-insulator substrate.

OCIS Codes: 230.5750, 230.5298, 220.4241

Owing to their ultrahigh-quality-factor whispering-gallery modes (WGMs) [1], microdisk and ring resonators are an essential building block in various applications ranging from optical communication [2] to nano-optomechanics [3], nonlinear optics [4], biochemical sensing [5] and so on. But such high quality ($Q$) factor is achieved at the cost of large footprint. As the disk (ring) radius reduces to the wavelength size, the bending loss from the total internal reflection significantly increases and degrades the $Q$ [6–8]. Photonic band gap (PBG) structures, such as circular grating resonators (CGRs) [9–11] and two-dimensional (2D) photonic crystal (PC) slabs [12, 13], provide the possibility to reduce the bending loss because light propagation is inhibited inside the band gap. However, the discontinuous structure of the CGRs is unsuitable for integrating with the widely used membrane fabrication technology, and the conventional triangular lattice PC needs sophisticated hole adjustment [14] to match the circular symmetry of the disk. Alternatively, circular photonic crystal (CPC) is considered as a very promising approach because it simultaneously possesses continuous slab structure and isotropic band gap [15, 16]. In this letter we present a 1.16-μm-radius Si disk embedded in a sunflower-type

CPC [16, 17] which allows us to overcome the radiation limit of the bare disks. We obtain an ultrahigh measured loaded optical $Q$ ($7.4 \times 10^5$), near an order of magnitude higher than that of a bare disk of similar size [6, 7, 18].

The proposed structure is shown in Fig. 1(a). In the sunflower-type CPC, a 220-nm-thick Si membrane is perforated by an array of holes with radius $r$, whose positions are determined by:

$$x_{N,l} = aN \cos\left(\frac{2l\pi}{6N}\right), \quad y_{N,l} = aN \sin\left(\frac{2l\pi}{6N}\right),$$

where $a$ is the lattice constant defined as the radial periodicity of the adjacent concentric rows, integer $N$ is the radial index, and $l = 1, 2, ..., 6N$ is the angular index of the holes. The innermost 3 concentric rows of holes are removed to form a C3 cavity to support whispering-gallery-like modes. A micron-sized disk with radius $R$, separated from the surrounding CPC by a gap $g$, is placed inside such a C3 cavity. Ideally, the confinement of the CPC can be further improved by radially chirping the hole array to maximally satisfy the phase-matching condition [17], but for $N \geq 3$ such chirping becomes negligible and therefore is not adopted in the proposed structure. The PBG effect of a perfect CPC lattice is investigated through three-dimensional (3D) finite-difference time-domain (FDTD) simulation, in which a TE-like polarized light wave excited by a Gaussian dipole source propagates through the CPC lattice and is probed on the other side. The simulated transmission spectrum for a perfect CPC lattice with a hole radius $r = 0.42a$ is plotted in Fig. 1(b) (dashed black curve), showing a band gap for normalized frequency ($a/\lambda$) ranging from 0.315 to 0.380 with at least 20 dB lower transmission than the pass band. When the disk is embedded, additional resonance peaks show up in the CPC band gap, each of which is labeled with its radial and azimuthal mode number ($m$, $n$) [solid red curve in Fig. 1(b)]. By comparing the mode profiles obtained from finite-element simulation [Fig. 1(c)], one concludes that the WGM (1, 9) possesses the best confinement. This is attributed to its small radial order number and the 9 azimuthal field maxima that match the 18 holes of the innermost concentric row of the surrounding CPC, thus making it a potential high-$Q$ mode.

Structure optimization is performed with 3D FDTD simulation to maximize the $Q$ factor of WGM (1, 9). First the lattice constant $a$ and the hole radius $r$ of the CPC structure are varied. Figures 2(a) and (b) show that the $Q$ first increases with both $a$ and $r$, but then drops drastically as $a$ exceeds 545 nm or $r$ exceeds 0.44$a$, when the resonance frequency moves away from the CPC band gap. Since a filling factor ($r/a$) of 0.44 is too large to get a robust CPC slab during fabrication, a value of 0.42 is a reasonable choice. Second, the gap $g$ between the surrounding CPC and the disk is optimized such that the confinement provided by the CPC can overcome its perturbation to the disk's WGMs, where an optimal value of $g$ = 110 nm is found to give the best $Q$ [Fig. 2(c)]. Another parameter that affects the $Q$ is the disk radius $R$, which determines the resonance frequency of the disk for a given band gap. Figure 2(d) shows that when $R$ is smaller than 1.12 µm, the $Q$ of the WGM (1, 9) from the sunflower–disk structure (circles) is lower than that of a bare disk of the same size (triangles), owing to the strong perturbation of the surrounding CPC. As the disk radius becomes larger than 1.12 µm, the confinement from the CPC band gap dominates and gives rise to a $Q$ as high as $10^7$, at least one order of magnitude higher than the bare disk. Note that once the band gap effect dominates, the $Q$ maintains at a relatively high level and does not vary much with the disk radius. Figure 2 has shown that the modal resonance frequency depends not only on the disk size but also on the CPC lattice parameters, indicating that the CPC lattice indeed affects the WGMs of the disk, which helps improving the modal $Q$.

The optimized geometry was fabricated on a standard silicon-on-insulator (SOI) substrate with 220-nm Si on 3-µm buried oxide layer. The pattern was first defined by e-beam lithography, and then transferred to the Si layer through chlorine-based inductively coupled plasma reactive ion etching, followed by a wet etching to remove the underlying buried oxide layer and form a freestanding CPC slab. The wet etching time was controlled such that the CPC slab was completely released while a $SiO_2$ supporting pedestal still remained beneath the disk. Figure 3(a) is an optical microscope image of a fabricated device, which shows the sunflower–disk structure in the middle and two grating couplers on the side for coupling light between the on-chip waveguide and optical fiber. A tilted-view scanning electron microscope (SEM) image in Fig. 3(b) shows a disk embedded in the CPC lattice with 20 concentric rows of holes. A line defect in the CPC serves as the coupling waveguide, which connects to the strip waveguide by a taper [Fig. 3(c)]. Owing to the residual compressive stress in the Si layer, the released CPC slab tends to buckle up, resulting in a vertical misalignment between the disk and the surrounding CPC. This misalignment would result in a reduced coupling efficiency between the waveguide and the disk and also a perturbation to the disk's modal field which degrades the $Q$. To relax the surface stress and eliminate the buckling effect, the CPC slab is connected to the bank with specially engineered clamps [Fig. 3(d)], yielding a good alignment between the disk and its surrounding CPC [Fig. 3(e)]. The fabricated devices were scaled up by about 2–3% from the design to compensate for the error introduced in fabrication processes and shift the resonance wavelength to around 1550 nm.

The devices are characterized by optical transmission measurements. Because of the in-line coupling configuration between the PC waveguide and the disk, light tunnels through the cavity only at the resonance frequencies, corresponding to the peaks in the transmission spectrum. Figure 4(a) plots the measured optical transmission spectrum of a fabricated device with disk radius $R$ = 1.16 μm and gap $g$ = 103 nm, showing two resonance peaks at 1536.5 nm and 1558.3 nm, which are identified by FDTD simulation as the WGMs (1, 9) and (2, 6), respectively. The optimized WGM (1, 9) targeted during the structure optimization exhibits a record high loaded $Q$ of $7.4 \times 10^5$, obtained by a Lorentzian fit of the zoomed-in spectrum [inset of Fig. 4(a)].

The measured $Q$ is affected by the coupling distance, defined as the number of holes ($K$) between the end of PC waveguide and the disk [Fig. 4(b)]. Figure 4(c) plots the measured transmission and the $Q$ as a function of $K$. Maximum transmission is found for $K$ = 3 with a $Q$ of $3.5 \times 10^5$, corresponding to an intrinsic $Q$ of around $7 \times 10^5$. It should be noted that the presence of the PC waveguide reduces the intrinsic $Q$ of the cavity, so increase of the coupling distance also results in an enhanced intrinsic $Q$.

In summary, we have demonstrated a novel approach for overcoming the radiation limit in micron-sized disks by embedding the disk into a CPC C3 cavity. The CPC structure matches the spatial distribution of the disk's WGMs at the boundary, and its band gap helps reducing the WGMs' bending loss. This implementation gives a simulated $Q$ as high as $10^7$ for the WGM (1, 9) of a micron-sized disk after 3D FDTD optimization, which is at least an order of magnitude higher than that of a bare disk with the same size. The design is experimentally verified by a record high loaded $Q$ of $7.4 \times 10^5$, measured from a device with a 1.16-μm-radius disk fabricated on a standard SOI substrate. The wavelength-sized disk radius and ultrahigh $Q$ make such sunflower–disk cavities an ideal candidate for applications in enhanced light–matter interaction, low-threshold lasing, high-frequency nano-optomechanical transduction, and so on.

This work is supported by DARPA/MTO ORCHID program through a grant from AFOSR, and National Science Foundation through a CAREER award. H. X. T. acknowledges support from a Packard Fellowship in Science and Engineering. The authors thank Michael Power and Dr. Michael Rooks for assistance in device fabrication.

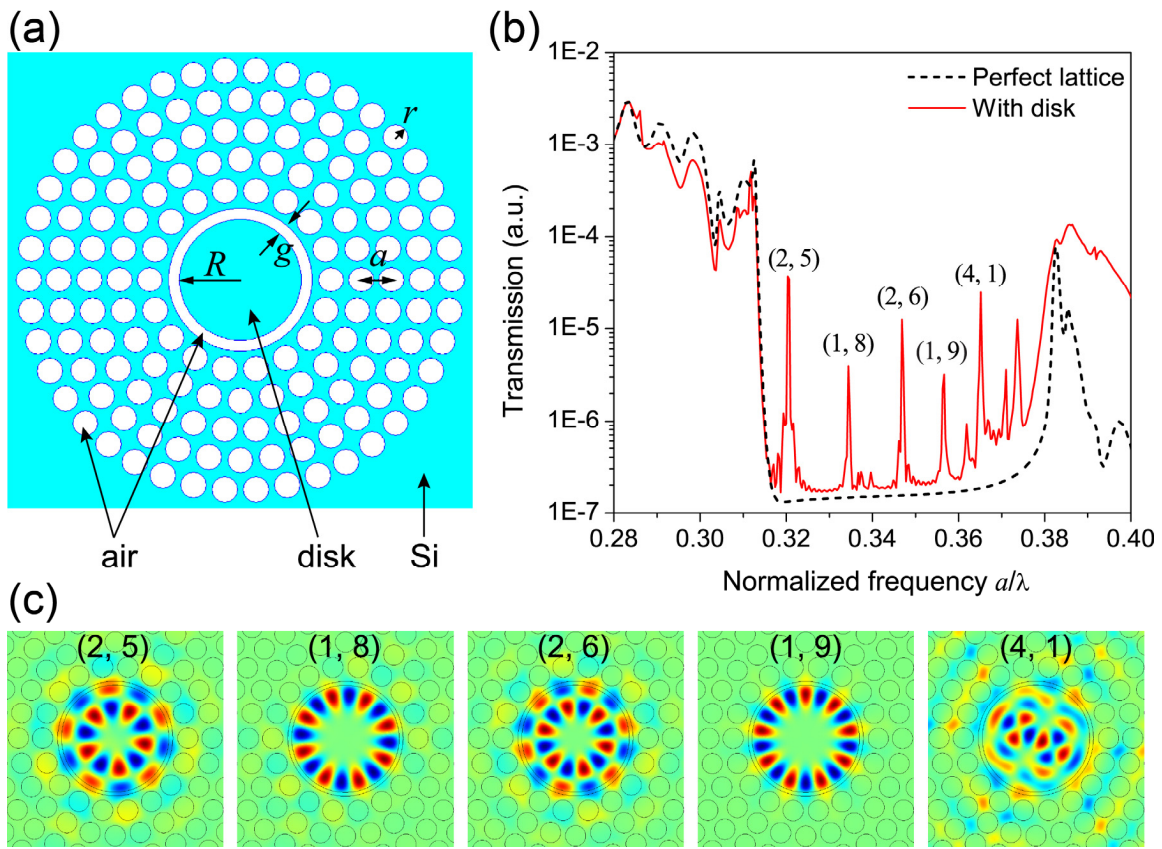

**Figure 1** (Color online) (a) Illustration of a sunflower-type circular photonic crystal (CPC) embedding a disk. (b) Simulated transmission spectrum of a perfect CPC lattice (dashed black line) and a CPC lattice embedding a disk (solid red line). Each resonance peak is labeled with its radial and azimuthal order number ($m$, $n$). (c) Finite-element-simulated mode profile ($H_z$ component) for the labeled modes in (b).

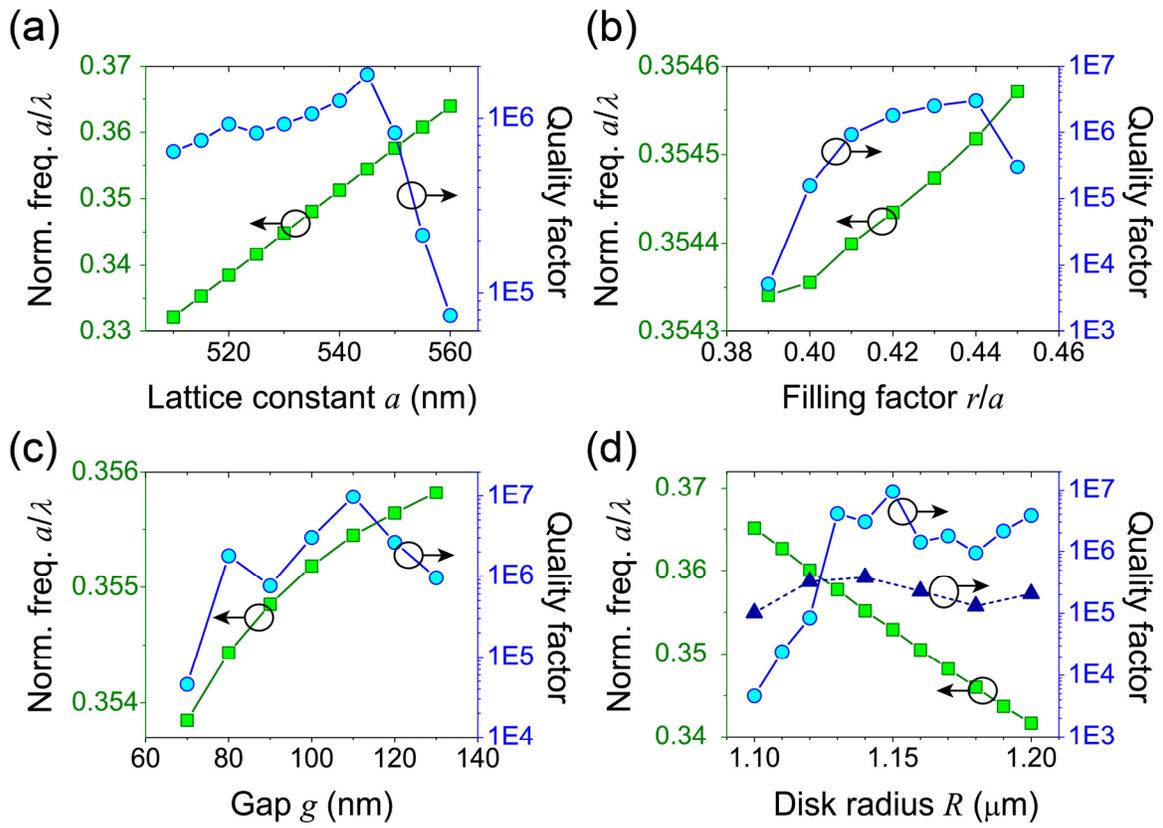

**Figure 2** (Color online) 3D-FDTD-simulated $Q$ factor (circles) and normalized resonance frequency $a/\lambda$ (squares) of whispering-gallery mode (1, 9) as a function of (a) lattice constant $a$, (b) filling factor $r/a$, (c) gap $g$ between the disk and the surrounding circular photonic crystal, and (d) disk radius $R$. The dashed line with triangles in (d) represents the simulated $Q$ factor for the same mode of a bare disk.

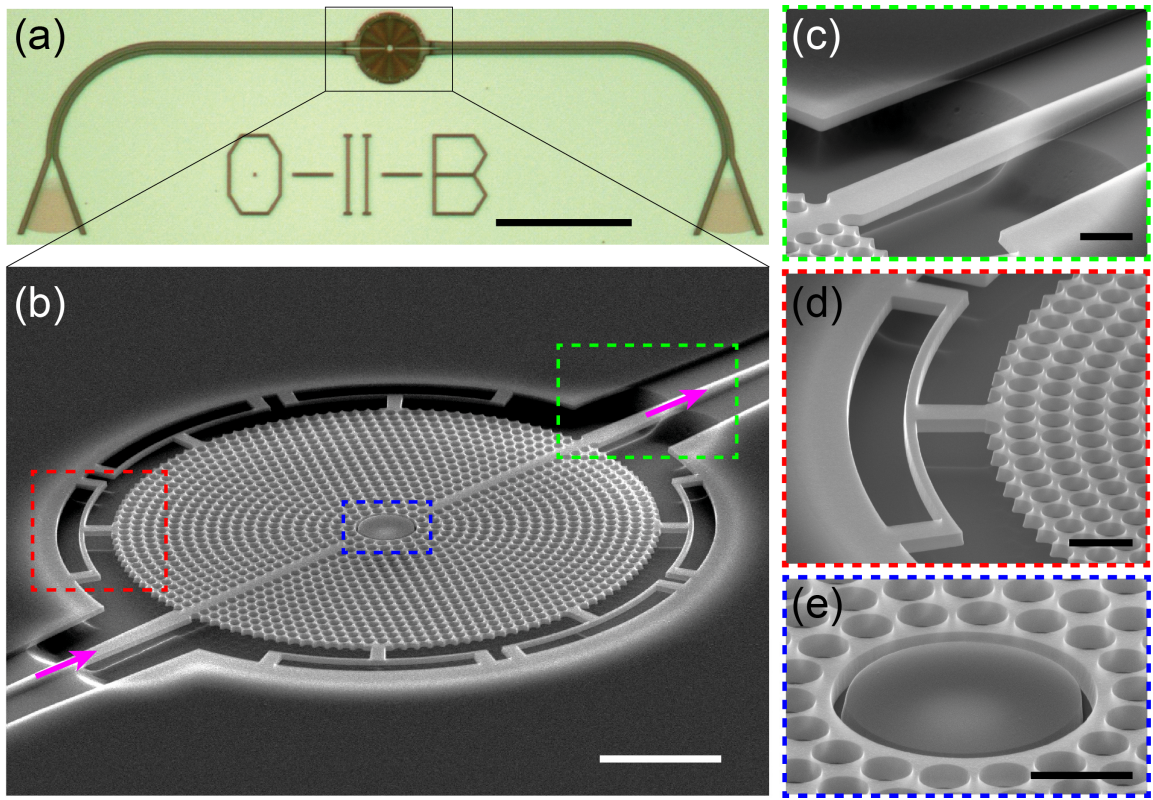

**Figure 3** (Color online) (a) Optical microscope image of a fabricated device. On the two sides are two grating couplers for coupling light onto and out of the chip. (b) Tilted-view scanning electron microscope (SEM) image of the sunflower-type circular photonic crystal with an embedded disk. Magenta arrows indicate the route of light propagation through the device. (c)–(e) Zoomed-in SEM image of a waveguide taper, a stress-releasing clamp, and the embedded disk, respectively. Scale bar is (a) 50 μm, (b) 5 μm, and (c)–(e) 1 μm, respectively.

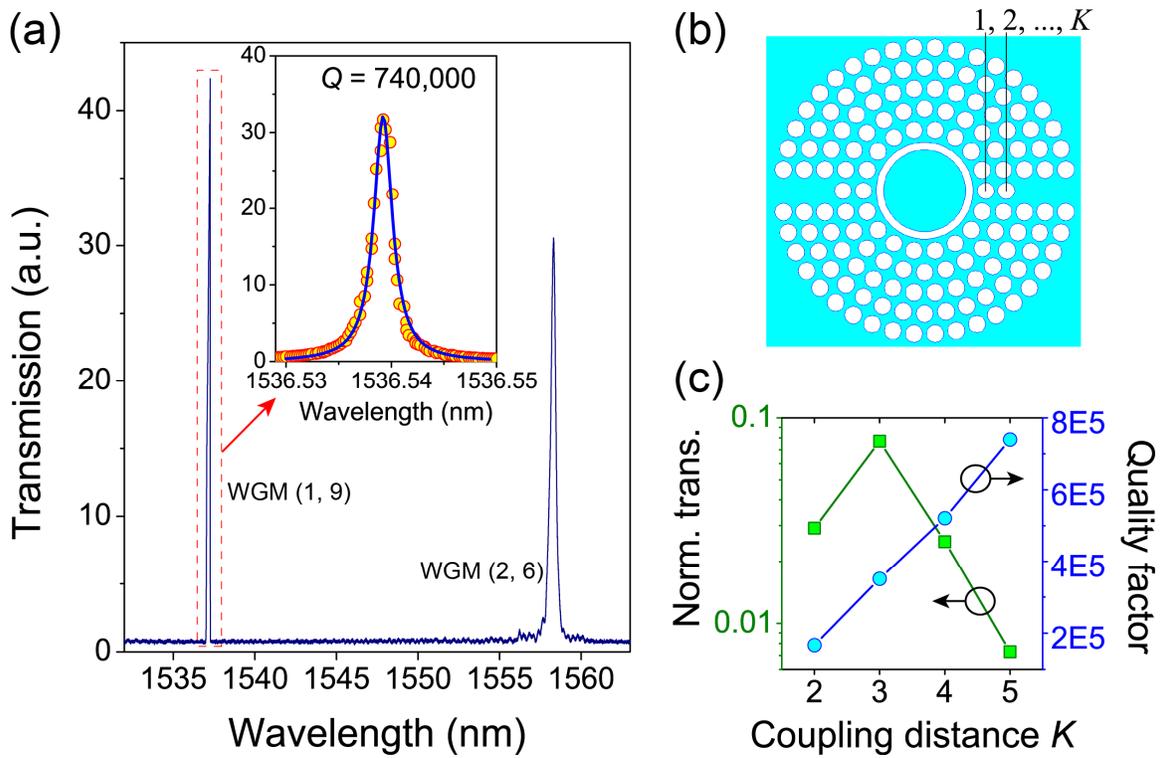

**Figure 4** (Color online) (a) Measured optical transmission of an optimized device. Left peak: WGM (1, 9); Right peak: WGM (2, 6). Inset: Lorentzian fit of the zoomed-in spectrum of the WGM (1, 9) revealing a loaded $Q$ of $7.4 \times 10^5$. (b) Illustration of the coupling distance, defined as the number of holes $K$ between the disk and the end of photonic crystal waveguide. (c) Measured $Q$ factor (circles) and normalized optical transmission (squares) as a function of the coupling distance $K$.